\documentclass[aps,pra,twocolumn,superscriptaddress,notitlepage,nofootinbib,longbibliography,floatfix,10pt]{revtex4-2}
\usepackage{bm}
\usepackage{amsmath,amssymb,amsthm}
\usepackage{mathrsfs}
\usepackage{physics}
\usepackage{dcolumn}
\usepackage{graphicx}
\usepackage{subfigure}
\usepackage[utf8]{inputenc}
\usepackage[T1]{fontenc}
\usepackage{url}
\usepackage{xcolor}
\usepackage[colorlinks=true,linkcolor=blue,anchorcolor=blue,citecolor=blue,urlcolor=magenta]{hyperref}
\usepackage{cleveref}
\crefname{equation}{Eq.}{Eqs.}
\crefrangeformat{equation}{Eqs.~(#3#1--#2#4)}
\newcommand{\ii}{\mathrm{i}}
\newcommand{\kB}{k_{\mathrm{B}}}
\newcommand{\veff}{V_{\mathrm{eff}}}
\newcommand{\sump}{\sideset{}{'}{\sum}}
\newcommand{\dgone}{\delta g_{1}}
\newcommand{\dgtwo}{\delta g_{2}}
\newcommand{\dg}{\delta g}

\begin{document}
\title{Universal quantum corrections of two-body correlation in a weakly interspecies interacting binary Bose mixture}
\author{Rui-Yan Chen}
\email{ruiyanchen@zjnu.edu.cn}
\author{Zhaoxin Liang}
\email{zhxliang@zjnu.edu.cn}
\author{Gao Xianlong}
\email{gaoxl@zjnu.edu.cn}
\affiliation{Department of Physics, Zhejiang Normal University, Jinhua 321004, China}
\date{\today}
\begin{abstract}
	We investigate universal quantum corrections to two-body correlations in a zero-temperature binary Bose mixture using the Cornwall-Jackiw-Tomboulis two-particle-irreducible effective action formalism. In the weak interspecies-coupling regime, a saddle-point treatment based on Hubbard-Stratonovich transformations can be combined with a two-loop expansion and a gapless Hartree-Fock correction, thereby preserving the Goldstone theorem and reducing the coupled two-component problem to two analytically solvable single-component theories. Within this framework, we derive the ground-state energy density as a low-density expansion in the gas parameter, together with the quantum depletion and chemical potentials. The results exhibit a simple mapping to the single-component case: the universal quantum corrections of the mixture are obtained by evaluating the known single-component series at an effective scattering length $a_{\sigma\sigma}-a_{12}$ for each species, where $a_{\sigma\sigma}$ and $a_{12}$ are the intra- and interspecies $s$-wave scattering lengths. This reproduces Petrov's equation of state at one-loop order in the weak-coupling limit and yields beyond-Lee-Huang-Yang corrections at two-loop order. We also analyze the role of mass imbalance, which enters the energy density through the exact rescaling factor $(1+m_{1}/m_{2})/2$.
\end{abstract}
\maketitle

\section{Introduction}
Ultracold quantum gases offer one of the cleanest laboratories for exploring how quantum fluctuations renormalize the properties of interacting many-body systems. Even in a homogeneous, weakly interacting Bose gas at zero temperature, the ground state is not a trivial condensate: zero-point fluctuations generate a series of universal quantum corrections to the equation of state (EOS). Starting with the pioneering works of Bogoliubov~\cite{bogoliubov1947}, Lee, Huang, and Yang~\cite{lhy1957}, and Wu~\cite{wu1959}, it was established that the equilibrium properties of a dilute Bose gas admit a systematic low-density expansion in powers of the gas parameter $\sqrt{n a^{3}}$, where $n$ is the particle density and $a$ the $s$-wave scattering length. The leading correction to the mean-field energy density, the celebrated Lee-Huang-Yang (LHY) term proportional to $n^{2}(na^{3})^{1/2}$, together with the accompanying quantum depletion $\propto n(na^{3})^{1/2}$, constitutes the so-called two-body-correlation universal quantum effect: at this order all microscopic details of the interatomic potential beyond $a$ are irrelevant~\cite{rmp76.599,braaten}. Beyond the LHY term, the next-order correction in the perturbative low-density expansion contains a logarithmic term of order $n^{2}(na^{3})\ln(na^{3})$ that arises from three-body correlations~\cite{wu1959}. The constant under the logarithm depends on a three-body coupling constant and was first evaluated by Braaten and Nieto~\cite{braaten}. Beyond that order the expansion becomes nonuniversal, being sensitive to the finite range of the interatomic potential~\cite{braaten,zhang2024} and, as established by Tan~\cite{tan2008}, to the three-body parameter.

A binary Bose mixture extends the physics of universal quantum corrections in a qualitatively richer direction. Following the first production of two overlapping condensates by sympathetic cooling~\cite{myatt1997}, binary condensates have been realized both in hyperfine-state mixtures of the same atomic species and in heteronuclear mixtures of distinct atomic species~\cite{thalhammer2008}. Their miscibility is governed by the celebrated criterion $g_{12}^{2}<g_{1}g_{2}$ for interspecies coupling~\cite{ho1996}. Thanks to Feshbach resonances, the interspecies $s$-wave scattering length $a_{12}$ is widely tunable, giving access to phenomena with no single-component counterpart. These include the miscible-immiscible phase transition when the interspecies repulsion exceeds the geometric mean of the intraspecies couplings~\cite{rakhimov2022}, and most remarkably, self-bound quantum droplets: when the interspecies attraction is strong enough to render the mean-field energy negative, mechanical collapse is stabilized by the repulsive LHY quantum fluctuations~\cite{petrov2015}, as confirmed experimentally in homonuclear~\cite{cabrera2018} and heteronuclear~\cite{derrico2019} mixtures. The droplet regime has stimulated intense theoretical activity, including consistent many-body treatments based on bosonic pairing that remove the loophole of a complex (softened) Bogoliubov spectrum in Petrov's theory~\cite{huhui}. However, most existing many-body theories of mixtures either remain at the one-loop (LHY) level or focus on strongly coupled regimes, e.g., phase separation under strong repulsion~\cite{rakhimov2022} and droplet formation under strong attraction~\cite{petrov2015,cabrera2018,derrico2019,huhui}. By contrast, the regime of \emph{weak} interspecies coupling, $\abs{a_{12}}\ll\abs{a_{11}},\abs{a_{22}}$, where the mixture stays miscible and the spectrum remains well defined for both attractive and repulsive $a_{12}$, has received much less attention, especially regarding quantum corrections \emph{beyond} the LHY term, and it is precisely in this regime that the universal quantum corrections of two-body correlation in a mixture should be established on a firm footing.

On the methodological side, the Cornwall-Jackiw-Tomboulis (CJT) two-particle-irreducible (2PI) effective action~\cite{CJT} provides a systematic framework for self-consistent quantum fluctuations, in which the Nambu-Goldstone theorem can be protected at the two-loop level by the gapless Hartree-Fock resummation scheme~\cite{dV}. Being a functional of the full (dressed) propagator, the 2PI effective action resums an infinite class of Feynman diagrams through the self-consistent dressing of the propagators, which is a nonperturbative partial resummation of the perturbative expansion. Its stationarity yields the Schwinger-Dyson equations. For a single-component Bose gas, this scheme has been shown to recover the full series of universal two-body-correlation corrections, i.e., the condensate fraction, energy density, chemical potential, and sound speed as expansions in the gas parameter~\cite{Song_2022}, and has recently been extended to the nonuniversal EOS with finite-range interactions, yielding analytical beyond-LHY corrections that are within reach of breathing-mode spectroscopy~\cite{zhang2024}. For binary mixtures, the CJT approach has been applied to finite-temperature phase transitions and symmetry-restoration phenomena~\cite{phat2009} and to the Casimir effect of confined dual Bose-Einstein condensates (BECs)~\cite{song2022aop}, yet a closed analytical two-loop EOS of a homogeneous mixture at zero temperature in the weakly coupled regime, i.e., the counterpart of the single-component low-density expansion, has remained absent. The technical bottleneck is that the interspecies interaction couples the two condensate sectors already at the level of the free propagator. The existing CJT treatments of binary mixtures~\cite{phat2009,song2022aop} effectively decouple the fluctuations of two species by adopting two independent $2\times2$ propagators, thereby omitting the mixed density-fluctuation modes of a full $4\times4$ treatment. The resulting coupled Schwinger-Dyson equations are difficult to solve analytically beyond the lowest order.

In this work, we remove this bottleneck by a different route, which, to the best of our knowledge, has not been applied to binary Bose mixtures before. We decouple the interspecies interaction via a Hubbard-Stratonovich transformation that introduces a real auxiliary field $\varphi$, with a real (imaginary) Gaussian weight for attractive (repulsive) interspecies coupling, and then treat the auxiliary field at the saddle-point level. The two species then decouple into two \emph{independent} single-component problems with renormalized intraspecies couplings $\delta g_{\sigma}=g_{\sigma}-g_{12}$ and shifted chemical potentials $\mu_{\sigma}+\hbar\varphi$. Each sector can be solved separately with the full machinery of the two-loop CJT effective action combined with the gapless Hartree-Fock correction~\cite{dV}, after which the auxiliary field is integrated out exactly by a Gaussian integral that restores the interspecies mean-field term. This construction is valid for both signs of $g_{12}$, preserves the Goldstone theorem in each sector, and reduces the coupled self-consistent problem to tractable single-component equations. The interspecies coupling is thereby included in a mixed and partial way, entering the fluctuations of each species through the renormalized couplings $\delta g_{\sigma}$ and the chemical-potential shifts, while the dynamical mixing of the two species' fluctuation modes is kept only at the static (saddle-point) level. The neglected dynamical part is of order $a_{12}/a$ and can be quantified by comparing with the exact one-loop results of Petrov~\cite{petrov2015}. The physical content of this scheme, which is the static analog of the random-phase approximation (RPA) of condensed-matter theory, is discussed in Sec.~\ref{sec:model}.

Within this scheme we study the zero-temperature condensed phase of the mixture, in which both components form BECs. The universal quantum corrections of two-body correlation considered here originate from the quantum fluctuations around the two condensates. Our main results are as follows. (i) At one-loop order, the EOS reproduces Petrov's result~\cite{petrov2015,huhui} in the symmetric limit and becomes exact as $a_{12}/a\to0$. (ii) At two-loop order, we obtain the beyond-LHY corrections as explicit series in the gas parameter, revealing a remarkably simple structure: the universal quantum corrections of the mixture are nothing but the single-component universal series evaluated at an \emph{effective} scattering length $\delta a_{\sigma}=a_{\sigma\sigma}-a_{12}$ for each species, while the direct interspecies contribution is entirely contained in the mean-field term. This ``single-component mapping'' encodes the two-body-correlation universal effect of the mixture in a compact and predictive way. (iii) We further analyze the effect of mass imbalance in heteronuclear mixtures, and find that, for symmetric densities and intraspecies scattering lengths, the entire energy density obeys $\mathcal{E}(n;m_{1},m_{2})=(1+m_{1}/m_{2})\mathcal{E}(n;m_{1},m_{1})/2$, an exact identity at both one- and two-loop orders.

The paper is organized as follows. In Sec.~\ref{sec:model}, we introduce the model and the Hubbard-Stratonovich decoupling scheme. Section~\ref{sec:energy} derives the effective potential and the EOS at one-loop and two-loop levels. In Sec.~\ref{sec:results}, we present and discuss our results, including comparisons with the existing one-loop (Petrov) theory and the analysis of mass imbalance. We conclude in Sec.~\ref{sec:conclusion}.

\section{Model}
\label{sec:model}
We consider a heteronuclear Bose-Bose mixture in three dimensions, in which both components form BECs. The system is described by the partition function
\begin{equation}
	\mathcal{Z}=\int\mathcal{D}\phi_{1}^{*}\mathcal{D}\phi_{1}\mathcal{D}\phi_{2}^{*}\mathcal{D}\phi_{2}\,e^{-S[\phi_{1}^{*},\phi_{1};\phi_{2}^{*},\phi_{2}]/\hbar}
\end{equation}
in coherent state path integral formalism~\cite{negele,stoof}, where the Euclidean action
\begin{align}
	S[\phi_{1}^{*},\phi_{1};\phi_{2}^{*},\phi_{2}]=\int_{x}\biggl\{&\phi_{1}^{*}(x)\left[\hbar\partial_{\tau}-\frac{\hbar^2\nabla^2}{2m_{1}}-\mu_{1}\right]\phi_{1}(x)\notag\\
	&+\phi_{2}^{*}(x)\left[\hbar\partial_{\tau}-\frac{\hbar^2\nabla^2}{2m_{2}}-\mu_{2}\right]\phi_{2}(x)\notag\\
	&+\frac{g_{1}}{2}\abs{\phi_{1}(x)}^4+\frac{g_{2}}{2}\abs{\phi_{2}(x)}^4\notag\\
	&+g_{12}\abs{\phi_{1}(x)}^2\abs{\phi_{2}(x)}^2\biggr\}.
\end{align}
Here, we use the standard notations $x=(\tau,\vb*{x})$ and $\int_{x}=\int dx=\int_{0}^{\hbar\beta}d\tau\int d^{D}\vb*{x}$, with $D=3$ and the inverse temperature $\beta\equiv1/(\kB T)$. The couplings $g_{1}=4\pi\hbar^2 a_{11}/m_{1}$ and $g_{2}=4\pi\hbar^2 a_{22}/m_{2}$, and $g_{12}=2\pi\hbar^2 a_{12}/m_{\mathrm{red}}$, with $a_{11}$, $a_{22}$, and $a_{12}$ being the corresponding $s$-wave scattering lengths, and with the reduced mass $m_{\mathrm{red}}=m_{1}m_{2}/(m_{1}+m_{2})$.

The interspecies interaction can be decoupled with Hubbard-Stratonovich transformations that introduce real auxiliary fields $\varphi$. When there is attraction between species, i.e., $g_{12}<0$, we employ the Gaussian integral
\begin{align}
	&\int\mathcal{D}\varphi\exp\biggl\{\int_{x}\biggl[\frac{1}{2}\varphi(x)\hbar g_{12}^{-1}\varphi(x)\notag\\
	&+\left(\abs{\phi_{1}(x)}^2+\abs{\phi_{2}(x)}^2\right)\varphi(x)\biggr]\biggr\}\notag\\
	=&\exp\biggl\{-\int_{x}\biggl[g_{12}\abs{\phi_{1}(x)}^2\abs{\phi_{2}(x)}^2\notag\\
	&+\frac{g_{12}}{2}\abs{\phi_{1}(x)}^4+\frac{g_{12}}{2}\abs{\phi_{2}(x)}^4\biggr]/\hbar\biggr\},
\end{align}
the theory then becomes
\begin{equation}
	\mathcal{Z}=\int\mathcal{D}\phi_{1}^{*}\mathcal{D}\phi_{1}\mathcal{D}\phi_{2}^{*}\mathcal{D}\phi_{2}\mathcal{D}\varphi\,e^{-S[\phi_{1}^{*},\phi_{1};\phi_{2}^{*},\phi_{2};\varphi]/\hbar},
\end{equation}
where the effective action
\begin{align}
	S[\phi_{1}^{*},\phi_{1};\phi_{2}^{*},\phi_{2};\varphi]=\int_{x}\biggl\{&\sum_{\sigma}\phi_{\sigma}^{*}(x)\left[\hbar\partial_{\tau}-\mu_{\sigma}\right]\phi_{\sigma}(x)\notag\\
	&-\sum_{\sigma}\frac{\hbar^2}{2m_{\sigma}}\phi_{\sigma}^{*}(x)\nabla^2\phi_{\sigma}(x)\notag\\
	&+\frac{\dgone}{2}\abs{\phi_{1}(x)}^4+\frac{\dgtwo}{2}\abs{\phi_{2}(x)}^4\notag\\
	&-\hbar\left(\abs{\phi_{1}(x)}^2+\abs{\phi_{2}(x)}^2\right)\varphi(x)\notag\\
	&-\frac{1}{2}\varphi(x)\hbar^2 g_{12}^{-1}\varphi(x)\biggr\}, \label{attracS}
\end{align}
with the sums running over $\sigma=1,2$, and $\dgone=g_{1}-g_{12}$, $\dgtwo=g_{2}-g_{12}$. When the interspecies interaction is repulsive, i.e., $g_{12}>0$, we use the Gaussian integral
\begin{align}
	&\int\mathcal{D}\varphi\exp\biggl\{-\int_{x}\biggl[\frac{1}{2}\varphi(x)\hbar g_{12}^{-1}\varphi(x)\notag\\
	&+\ii\left(\abs{\phi_{1}(x)}^2+\abs{\phi_{2}(x)}^2\right)\varphi(x)\biggr]\biggr\}\notag\\
	=&\exp\biggl\{-\int_{x}\biggl[g_{12}\abs{\phi_{1}(x)}^2\abs{\phi_{2}(x)}^2\notag\\
	&+\frac{g_{12}}{2}\abs{\phi_{1}(x)}^4+\frac{g_{12}}{2}\abs{\phi_{2}(x)}^4\biggr]/\hbar\biggr\},
\end{align}
which leads to the effective action
\begin{align}
	S[\phi_{1}^{*},\phi_{1};\phi_{2}^{*},\phi_{2};\varphi]=\int_{x}\biggl\{&\sum_{\sigma}\phi_{\sigma}^{*}(x)\left[\hbar\partial_{\tau}-\mu_{\sigma}\right]\phi_{\sigma}(x)\notag\\
	&-\sum_{\sigma}\frac{\hbar^2}{2m_{\sigma}}\phi_{\sigma}^{*}(x)\nabla^2\phi_{\sigma}(x)\notag\\
	&+\frac{\dgone}{2}\abs{\phi_{1}(x)}^4+\frac{\dgtwo}{2}\abs{\phi_{2}(x)}^4\notag\\
	&+\ii\hbar\left(\abs{\phi_{1}(x)}^2+\abs{\phi_{2}(x)}^2\right)\varphi(x)\notag\\
	&+\frac{1}{2}\varphi(x)\hbar^2 g_{12}^{-1}\varphi(x)\biggr\}. \label{repulS}
\end{align}
Next, we adopt the saddle-point approximation and take a uniform solution $\varphi(x)=\varphi$. In this way, we can first perform integration on $\{\phi_{1}^{*},\phi_{1}\}$ and $\{\phi_{2}^{*},\phi_{2}\}$ separately. We expect this approximate treatment to be reliable, at least for weak interspecies coupling, i.e., $\abs{a_{12}}\ll\abs{a_{11}},\abs{a_{22}}$, and it makes the inclusion of higher-order quantum fluctuations more tractable.

To appreciate the advantage of this route, recall that a direct treatment of the mixture starts with the $4\times4$ free inverse propagator $G_{0}^{-1}(k)$, whose off-diagonal blocks, proportional to $g_{12}v_{1}v_{2}$ (where $v_{1},v_{2}$ denote the condensate amplitudes of the two components), couple the density fluctuations of the two components. Existing CJT calculations for binary mixtures~\cite{phat2009,song2022aop} instead take two independent $2\times2$ propagators and thereby omit these mixed fluctuation modes already at the free-propagator level. In our scheme, the interspecies coupling is absorbed into the auxiliary field: the saddle point implements its static (Hartree) level, and the dynamical part of the mixed modes enters only at order $a_{12}/a$, being systematically neglected in the weakly coupled regime.

Conceptually, this construction is the static analog of the random-phase approximation (RPA) familiar from condensed-matter many-body theory. A complete RPA treatment~\cite{nagaosa} would integrate the auxiliary field as a momentum-dependent Gaussian fluctuation, thereby resumming the density-density (ring) diagrams of the interspecies channel. Here, the saddle-point approximation freezes the auxiliary field at its uniform classical value, implementing the Hartree mean field of the interspecies channel, and the residual uniform Gaussian fluctuations, which we integrate out exactly, restore the mean field interspecies energy. This static level is justified in the weakly coupled regime, where the dynamical, momentum-dependent corrections are suppressed by the powers of $a_{12}/a$, the small parameter controlling our expansion. The scheme is non-trivial in two respects. First, it converts the coupled two-species self-consistent problem into two independent single-component problems, each of which admits an analytical two-loop solution with the Goldstone theorem protected, so that the full low-density expansion of the mixture EOS is obtained in closed form. Second, the final expressions apply uniformly to attractive and repulsive interspecies interactions: the sign of $g_{12}$ enters the Hubbard-Stratonovich decoupling only through the choice of a real or imaginary Gaussian weight, and the subsequent calculation, including the saddle-point treatment and the two-loop expansion, proceeds identically in both cases.

\section{Loop expansion, energy density, and quantum depletion}
\label{sec:energy}
Within the saddle-point approximation for weak interspecies interactions, we perform self-consistent calculations by a loop expansion of the 2PI effective action, and obtain the energy density of the system when macroscopic condensation occurs at zero temperature.
\subsection{One-loop level}
For the case of interspecies attraction, according to Eq.~\eqref{attracS} (the repulsive case $g_{12}>0$ follows analogously from Eq.~\eqref{repulS}, with an imaginary Gaussian weight for the auxiliary field, and yields identical final expressions), in the saddle-point approximation,
\begin{align}
	S\simeq&\hbar\beta L^D\biggl[-\left(\mu_{1}+\hbar\varphi\right)v_{1}^2+\frac{\dgone}{2}v_{1}^4-\left(\mu_{2}+\hbar\varphi\right)v_{2}^2+\frac{\dgtwo}{2}v_{2}^4\notag\\
	&-\frac{1}{2}\varphi\hbar^2 g_{12}^{-1}\varphi\biggr]\notag\\
	&+\frac{1}{2}\sump_{k}\vb*{\eta}^{\top}(-k)G_{0}^{-1}(k)\vb*{\eta}(k)\notag\\
	&+\int_{x}\biggl\{\frac{\dgone}{\sqrt{2}}v_{1}\left(\eta_{1}^3+\eta_{1}\eta_{2}^2\right)+\frac{\dgone}{8}\left(\eta_{1}^2+\eta_{2}^2\right)^2\notag\\
	&+\frac{\dgtwo}{\sqrt{2}}v_{2}\left(\eta_{3}^3+\eta_{3}\eta_{4}^2\right)+\frac{\dgtwo}{8}\left(\eta_{3}^2+\eta_{4}^2\right)^2\biggr\}.
\end{align}
Here, we have separated the field into its coherent and fluctuating components in real field formalism, i.e., $\phi_{1}(x)=v_{1}+\left[\eta_{1}(x)+\ii\eta_{2}(x)\right]/\sqrt{2}$, $\phi_{2}(x)=v_{2}+\left[\eta_{3}(x)+\ii\eta_{4}(x)\right]/\sqrt{2}$, and $\vb*{\eta}=\mqty(\eta_{1}&\eta_{2}&\eta_{3}&\eta_{4})^{\top}$. The notation $\sump_{k}=\frac{1}{\hbar\beta}\sum_{\omega_{n}}\frac{1}{L^D}\sum_{\vb*{k}}=\frac{1}{\hbar\beta L^D}\sum_{k}$ arises from the Fourier transforms, and $\omega_{n}=2\pi n/(\hbar\beta)$ are the bosonic Matsubara frequencies, $k=(\omega_{n},\vb*{k})$. The inverse propagator
\begin{equation}
	G_{0}^{-1}(k)=\mqty(A_{1}&\hbar\omega_{n}&0&0\\
	-\hbar\omega_{n}&A_{2}&0&0\\
	0&0&A_{3}&\hbar\omega_{n}\\
	0&0&-\hbar\omega_{n}&A_{4}),
\end{equation}
where
\begin{align}
	A_{1}=&\frac{\hbar^2\vb*{k}^2}{2m_{1}}-\mu_{1}-\hbar\varphi+3\dgone v_{1}^2,\\
	A_{2}=&\frac{\hbar^2\vb*{k}^2}{2m_{1}}-\mu_{1}-\hbar\varphi+\dgone v_{1}^2,\\
	A_{3}=&\frac{\hbar^2\vb*{k}^2}{2m_{2}}-\mu_{2}-\hbar\varphi+3\dgtwo v_{2}^2,\\
	A_{4}=&\frac{\hbar^2\vb*{k}^2}{2m_{2}}-\mu_{2}-\hbar\varphi+\dgtwo v_{2}^2.
\end{align}

The partial one-loop effective potential
\begin{align}
	\veff[v]=&-\left(\mu_{1}+\hbar\varphi\right)v_{1}^2+\frac{\dgone}{2}v_{1}^4-\left(\mu_{2}+\hbar\varphi\right)v_{2}^2+\frac{\dgtwo}{2}v_{2}^4\notag\\
	&+\frac{\hbar}{2}\sump_{k}\tr\ln(G_{0}^{-1}[v](k)/\hbar),
\end{align}
after summing over Matsubara frequencies,
\begin{equation}
	\frac{\hbar}{2}\sump_{k}\tr\ln(G_{0}^{-1}/\hbar)=\frac{1}{2}\frac{1}{L^D}\sum_{\vb*{k}}\left[\hbar\omega_{1\vb*{k}}+\hbar\omega_{2\vb*{k}}\right]
\end{equation}
at zero temperature, where $\hbar\omega_{1\vb*{k}}=\sqrt{A_{1}A_{2}}$, $\hbar\omega_{2\vb*{k}}=\sqrt{A_{3}A_{4}}$. In the tree approximation, i.e., neglecting the quantum fluctuations, by taking the derivative of the condensate thermodynamic potential with respect to $v_{1}$ and $v_{2}$, we get
\begin{align}
	-\mu_{1}-\hbar\varphi+\dgone v_{1}^2=&0,\\
	-\mu_{2}-\hbar\varphi+\dgtwo v_{2}^2=&0,
\end{align}
and the well-known gapless Bogoliubov spectra
\begin{align}
	\hbar\omega_{1\vb*{k}}=&\sqrt{\frac{\hbar^2\vb*{k}^2}{2m_{1}}\left(\frac{\hbar^2\vb*{k}^2}{2m_{1}}+M_{1}\right)},\\
	\hbar\omega_{2\vb*{k}}=&\sqrt{\frac{\hbar^2\vb*{k}^2}{2m_{2}}\left(\frac{\hbar^2\vb*{k}^2}{2m_{2}}+M_{2}\right)},
\end{align}
where $M_{1}=2\dgone v_{1}^2$, $M_{2}=2\dgtwo v_{2}^2$. In the thermodynamic limit (particle number $N\to\infty$, system size $L\to\infty$, particle density $n=N/V\to\text{finite}$), $\frac{1}{L^D}\sum_{\vb*{k}}\to\int\frac{d^D\vb*{k}}{(2\pi)^D}$, after integration and regularization, we can obtain
\begin{align}
	\veff[v]=&-\left(\mu_{1}+\hbar\varphi\right)v_{1}^2+\frac{\dgone}{2}v_{1}^4-\left(\mu_{2}+\hbar\varphi\right)v_{2}^2+\frac{\dgtwo}{2}v_{2}^4\notag\\
	&+\frac{\sqrt{2}m_{1}^{3/2}M_{1}^{5/2}}{15\pi^2\hbar^3}+\frac{\sqrt{2}m_{2}^{3/2}M_{2}^{5/2}}{15\pi^2\hbar^3}.
\end{align}
Thus,
\begin{equation}
	\mathcal{Z}=\int d\varphi\exp\left\{\beta L^D\left(\frac{1}{2}\varphi\hbar^2 g_{12}^{-1}\varphi-\veff[v]\right)\right\},
\end{equation}
and we can perform a Gaussian integral on the auxiliary field $\varphi$, which yields
\begin{align}
	\mathcal{Z}=\exp\biggl\{&-\beta L^D\biggl(-\mu_{1}v_{1}^2-\mu_{2}v_{2}^2+\frac{g_{1}}{2}v_{1}^4+\frac{g_{2}}{2}v_{2}^4+g_{12}v_{1}^2v_{2}^2\notag\\
	&+\frac{\sqrt{2}m_{1}^{3/2}M_{1}^{5/2}}{15\pi^2\hbar^3}+\frac{\sqrt{2}m_{2}^{3/2}M_{2}^{5/2}}{15\pi^2\hbar^3}\biggr)\biggr\}.
\end{align}

According to the standard thermodynamic definitions, $\mathcal{Z}=\exp(-\beta L^D\veff)$ and $\veff=-P(\mu)$, which is related to energy density by the Legendre transformation $\mathcal{E}(n)=\mu n-P(\mu)$, $n=\partial P/\partial\mu$. This yields
\begin{align}
	\mathcal{E}=&\frac{g_{1}}{2}n_{1}^2+\frac{g_{2}}{2}n_{2}^2+g_{12}n_{1}n_{2}\notag\\
	&+\frac{8m_{1}^{3/2}\dgone^{5/2}}{15\pi^2\hbar^3}n_{1}^{5/2}+\frac{8m_{2}^{3/2}\dgtwo^{5/2}}{15\pi^2\hbar^3}n_{2}^{5/2},
\end{align}
with which we take $n_{1}=v_{1}^2$, $n_{2}=v_{2}^2$. When we focus on the idealized case of $m_{1}=m_{2}=m$, $a_{11}=a_{22}=a$ and $n_{1}=n_{2}=n/2$, we find that
\begin{align}
	\mathcal{E}=&\frac{\pi\hbar^2}{m}(a+a_{12})n^2\notag\\
	&+\frac{64\sqrt{2\pi}\hbar^2}{15m}a^{5/2}\left(1-\frac{a_{12}}{a}\right)^{5/2}n^{5/2}.
\end{align}
For comparison, Petrov's one-loop result~\cite{huhui,petrov2015} reads
\begin{equation}
	\mathcal{E}=\frac{\pi\hbar^2}{m}(a+a_{12})n^2+\frac{32\sqrt{2\pi}\hbar^2}{15m}a^{5/2}\mathcal{F}(\frac{a_{12}}{a})n^{5/2},
\end{equation}
where $\mathcal{F}(\alpha)=(1+\alpha)^{5/2}+(1-\alpha)^{5/2}$ becomes complex in the droplet phase $a+a_{12}<0$. In the weak-coupling regime $\abs{a_{12}/a}\ll1$ the two expressions coincide up to corrections of order $a_{12}/a$.

\subsection{Two-loop approximation}
Let us now further consider the calculation of effective potential in the two-loop approximation. Since we neglect the setting sun diagrams and only include the double bubble diagrams in two-loop skeleton diagrams, the corresponding contribution, known as the Luttinger-Ward functional in the condensed-matter literature~\cite{luttinger1960} (equivalently, the $\Phi$ functional of the $\Phi$-derivable scheme~\cite{rmp76.599}, i.e., of the CJT 2PI effective action~\cite{CJT}), reads
\begin{align}
	V_{2}[G]=&\frac{3\dgone}{8}\left(Q_{11}^2+Q_{22}^2\right)+\frac{\dgone}{4}Q_{11}Q_{22}\notag\\
	&+\frac{3\dgtwo}{8}\left(Q_{33}^2+Q_{44}^2\right)+\frac{\dgtwo}{4}Q_{33}Q_{44},
\end{align}
where $Q_{ij}=\sump_{k}\hbar G_{ij}$, $i,j=1,2,3,4$. To protect the Nambu-Goldstone theorem when taking into account the field fluctuations, a gapless Hartree-Fock approximation introduces a phenomenological symmetry-restoring correction~\cite{hugenholtz1959,Song_2022,dV}
\begin{align}
	\Delta V=&-\frac{\dgone}{4}\left(Q_{11}^2+Q_{22}^2\right)+\frac{\dgone}{2}Q_{11}Q_{22}\notag\\
	&-\frac{\dgtwo}{4}\left(Q_{33}^2+Q_{44}^2\right)+\frac{\dgtwo}{2}Q_{33}Q_{44}
\end{align}
to the $\Phi$-derivable scheme. Therefore, the effective potential in the two-loop approximation takes the form
\begin{align}
	\veff[v,G]=&-\left(\mu_{1}+\hbar\varphi\right)v_{1}^2+\frac{\dgone}{2}v_{1}^4-\left(\mu_{2}+\hbar\varphi\right)v_{2}^2+\frac{\dgtwo}{2}v_{2}^4\notag\\
	&+\frac{1}{2}\sump_{k}\tr\left[\ln G^{-1}(k)+G_{0}^{-1}[v]G(k)\right]\notag\\
	&+V_{2}[v,G]+\Delta V,
\end{align}
which gives
\begin{align}
	-(\mu_{1}+\hbar\varphi)+\dgone v_{1}^2+\frac{3\dgone}{2}Q_{11}+\frac{\dgone}{2}Q_{22}=0,\\
	-(\mu_{2}+\hbar\varphi)+\dgtwo v_{2}^2+\frac{3\dgtwo}{2}Q_{33}+\frac{\dgtwo}{2}Q_{44}=0,
\end{align}
by taking the variation of the effective potential with respect to $v_{1}$ and $v_{2}$. Within the 2PI formalism, the Schwinger-Dyson equations follow equivalently from the stationarity of the effective potential with respect to the full propagator, $\delta\veff[v,G]/\delta G=0$, which yields $G^{-1}=G_{0}^{-1}-\Sigma$ with
\begin{align}
	-\Sigma_{11}=&\frac{\dgone}{2}Q_{11}+\frac{3\dgone}{2}Q_{22},\label{selfe1}\\
	-\Sigma_{22}=&\frac{\dgone}{2}Q_{22}+\frac{3\dgone}{2}Q_{11},\label{selfe2}\\
	-\Sigma_{33}=&\frac{\dgtwo}{2}Q_{33}+\frac{3\dgtwo}{2}Q_{44},\label{selfe3}\\
	-\Sigma_{44}=&\frac{\dgtwo}{2}Q_{44}+\frac{3\dgtwo}{2}Q_{33},\label{selfe4}
\end{align}
we can obtain the gap equations
\begin{align}
	-(\mu_{1}+\hbar\varphi)+\dgone v_{1}^2-\Sigma_{22}=0,\label{gap1}\\
	-(\mu_{2}+\hbar\varphi)+\dgtwo v_{2}^2-\Sigma_{44}=0,\label{gap2}
\end{align}
and the full propagator
\begin{equation}
	G^{-1}(k)=\mqty(\frac{\hbar^2\vb*{k}^2}{2m_{1}}+M_{1}&\hbar\omega_{n}&0&0\\
	-\hbar\omega_{n}&\frac{\hbar^2\vb*{k}^2}{2m_{1}}&0&0\\
	0&0&\frac{\hbar^2\vb*{k}^2}{2m_{2}}+M_{2}&\hbar\omega_{n}\\
	0&0&-\hbar\omega_{n}&\frac{\hbar^2\vb*{k}^2}{2m_{2}}),
\end{equation}
where
\begin{align}
	M_{1}=-(\mu_{1}+\hbar\varphi)+3\dgone v_{1}^2-\Sigma_{11},\label{DS1}\\
	M_{2}=-(\mu_{2}+\hbar\varphi)+3\dgtwo v_{2}^2-\Sigma_{33}.\label{DS2}
\end{align}
By solving the poles of the Green's function, we obtain the two Bogoliubov spectra
\begin{align}
	\hbar\omega_{1\vb*{k}}=&\sqrt{\frac{\hbar^2\vb*{k}^2}{2m_{1}}\left(\frac{\hbar^2\vb*{k}^2}{2m_{1}}+M_{1}\right)},\\
	\hbar\omega_{2\vb*{k}}=&\sqrt{\frac{\hbar^2\vb*{k}^2}{2m_{2}}\left(\frac{\hbar^2\vb*{k}^2}{2m_{2}}+M_{2}\right)}.
\end{align}
At zero temperature, after integrating and regularizing in the continuum limit, we get
\begin{align}
	Q_{11}=&\frac{\sqrt{2}m_{1}^{3/2}M_{1}^{3/2}}{3\pi^2\hbar^3},\\
	Q_{22}=&-\frac{\sqrt{2}m_{1}^{3/2}M_{1}^{3/2}}{6\pi^2\hbar^3},\\
	Q_{33}=&\frac{\sqrt{2}m_{2}^{3/2}M_{2}^{3/2}}{3\pi^2\hbar^3},\\
	Q_{44}=&-\frac{\sqrt{2}m_{2}^{3/2}M_{2}^{3/2}}{6\pi^2\hbar^3},
\end{align}
and
\begin{widetext}
	\begin{align}
		\frac{\hbar}{2}\sump_{k}\tr(G_{0}^{-1}[v]G)=&\frac{1}{2}\left[\left(-(\mu_{1}+\hbar\varphi)+3\dgone v_{1}^2-M_{1}\right)Q_{11}+\left(-(\mu_{1}+\hbar\varphi)+\dgone v_{1}^2\right)Q_{22}\right]\notag\\
		&+\frac{1}{2}\left[\left(-(\mu_{2}+\hbar\varphi)+3\dgtwo v_{2}^2-M_{2}\right)Q_{33}+\left(-(\mu_{2}+\hbar\varphi)+\dgtwo v_{2}^2\right)Q_{44}\right].
	\end{align}
\end{widetext}
Starting from the partition function
\begin{equation}
	\mathcal{Z}=\int d\varphi\exp\left\{\beta L^D\left(\frac{1}{2}\varphi\hbar^2 g_{12}^{-1}\varphi-\veff[v,G]\right)\right\}
\end{equation}
and integrating over $\varphi$, we find
\begin{widetext}
	\begin{align}
		-P(\mu)=\veff=&-\mu_{1}v_{1}^2-\mu_{2}v_{2}^2+\frac{\dgone}{2}v_{1}^4+\frac{\dgtwo}{2}v_{2}^4+\frac{\sqrt{2}m_{1}^{3/2}M_{1}^{5/2}}{15\pi^2\hbar^3}+\frac{\sqrt{2}m_{2}^{3/2}M_{2}^{5/2}}{15\pi^2\hbar^3}\notag\\
		&+\frac{\dgone}{8}\left(Q_{11}^2+Q_{22}^2\right)+\frac{3\dgone}{4}Q_{11}Q_{22}+\frac{\dgtwo}{8}\left(Q_{33}^2+Q_{44}^2\right)+\frac{3\dgtwo}{4}Q_{33}Q_{44}\notag\\
		&+\frac{1}{2}\left[\left(-\mu_{1}+3\dgone v_{1}^2-M_{1}\right)Q_{11}+\left(-\mu_{1}+\dgone v_{1}^2\right)Q_{22}\right]\notag\\
		&+\frac{1}{2}\left[\left(-\mu_{2}+3\dgtwo v_{2}^2-M_{2}\right)Q_{33}+\left(-\mu_{2}+\dgtwo v_{2}^2\right)Q_{44}\right]\notag\\
		&+\frac{g_{12}}{2}\left[v_{1}^2+\frac{1}{2}\left(Q_{11}+Q_{22}\right)+v_{2}^2+\frac{1}{2}\left(Q_{33}+Q_{44}\right)\right]^2. \label{P}
	\end{align}
\end{widetext}

The particle densities are determined by
\begin{align}
	n_{1}=&\pdv{P}{\mu_{1}}=v_{1}^2+\frac{1}{2}\left(Q_{11}+Q_{22}\right),\label{n1}\\
	n_{2}=&\pdv{P}{\mu_{2}}=v_{2}^2+\frac{1}{2}\left(Q_{33}+Q_{44}\right),\label{n2}
\end{align}
which can be combined with \cref{selfe1,selfe2,selfe3,selfe4,gap1,gap2} and \eqref{DS1}, \eqref{DS2}, and leads to the self-consistent equations
\begin{align}
	M_{1}=&2\dgone n_{1}-2\dgone Q_{11},\label{M1}\\
	M_{2}=&2\dgtwo n_{2}-2\dgtwo Q_{33}.\label{M2}
\end{align}
We can solve Eqs.~\eqref{M1}, \eqref{M2} and expand the solutions in the gas parameters $\sqrt{n_{1}(a_{11}-a_{12})^3}$ and $\sqrt{n_{2}(a_{22}-a_{12})^3}$, respectively, for a dilute (weakly interacting) mixture, which yields
\begin{widetext}
	\begin{align}
		x_{1}\equiv&\sqrt{M_{1}}=\sqrt{2\dgone n_{1}}\left[1-\frac{16\sqrt{n_{1}(a_{11}-a_{12})^3}}{3\sqrt{\pi}}\left(1-\frac{40\sqrt{n_{1}(a_{11}-a_{12})^3}}{3\sqrt{\pi}}\right)+\mathcal{O}\left(\left(n_{1}(a_{11}-a_{12})^3\right)^{3/2}\right)\right],\\
		x_{2}\equiv&\sqrt{M_{2}}=\sqrt{2\dgtwo n_{2}}\left[1-\frac{16\sqrt{n_{2}(a_{22}-a_{12})^3}}{3\sqrt{\pi}}\left(1-\frac{40\sqrt{n_{2}(a_{22}-a_{12})^3}}{3\sqrt{\pi}}\right)+\mathcal{O}\left(\left(n_{2}(a_{22}-a_{12})^3\right)^{3/2}\right)\right].
	\end{align}
\end{widetext}
Using \cref{P,n1,n2}, we get the EOS
\begin{widetext}
\begin{align}
	\mathcal{E}(n)=&\mu n-P(\mu)\notag\\
	=&\frac{\dgone}{2}v_{1}^4+\frac{\dgtwo}{2}v_{2}^4\notag\\
	&+\frac{\sqrt{2}m_{1}^{3/2}M_{1}^{5/2}}{15\pi^2\hbar^3}+\frac{\sqrt{2}m_{2}^{3/2}M_{2}^{5/2}}{15\pi^2\hbar^3}\notag\\
	&+\frac{\dgone}{8}\left(Q_{11}^2+Q_{22}^2\right)+\frac{3\dgone}{4}Q_{11}Q_{22}\notag\\
	&+\frac{\dgtwo}{8}\left(Q_{33}^2+Q_{44}^2\right)+\frac{3\dgtwo}{4}Q_{33}Q_{44}\notag\\
	&+\frac{1}{2}\left[\left(3\dgone v_{1}^2-M_{1}\right)Q_{11}+\left(\dgone v_{1}^2\right)Q_{22}\right]\notag\\
	&+\frac{1}{2}\left[\left(3\dgtwo v_{2}^2-M_{2}\right)Q_{33}+\left(\dgtwo v_{2}^2\right)Q_{44}\right]\notag\\
	&+\frac{g_{12}}{2}\left(n_{1}+n_{2}\right)^2,
\end{align}
\end{widetext}
then we take $v_{1}^2=n_{1}-(Q_{11}+Q_{22})/2$ and $v_{2}^2=n_{2}-(Q_{33}+Q_{44})/2$ from Eqs.~\eqref{n1} and \eqref{n2}, so that a self-consistent universal energy density expressed in terms of the particle densities and the microscopic interaction parameters is obtained. In addition, the analytical expression of the quantum depletion
\begin{align}
	n_{\mathrm{ex}}&=n_{1\mathrm{ex}}+n_{2\mathrm{ex}}=(n_{1}-v_{1}^2)+(n_{2}-v_{2}^2)\notag\\
	&=\frac{1}{2}\left(Q_{11}+Q_{22}+Q_{33}+Q_{44}\right)
\end{align}
can be derived by substituting the solutions for $x_{1}$ and $x_{2}$.

\section{Results}
\label{sec:results}
\begin{figure}
	\centering
	\includegraphics[width=0.45\textwidth]{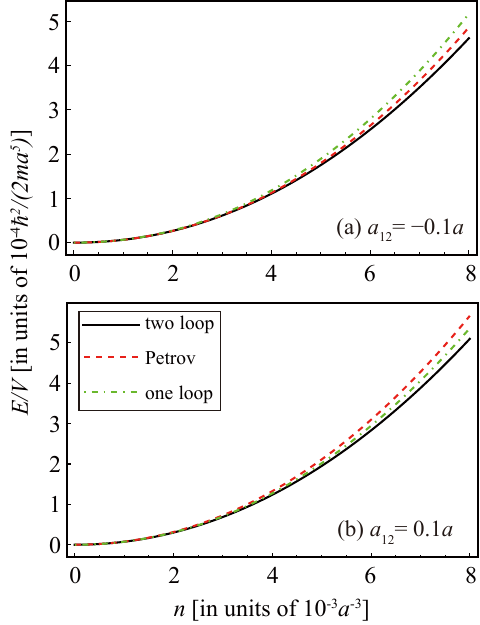}
	\caption{Energy density $\mathcal{E}$ as a function of the total particle density $n$ for (a) $a_{12}=-0.1a$ and (b) $a_{12}=0.1a$, with $m_{1}=m_{2}=m$, $a_{11}=a_{22}=a$, and $n_{1}=n_{2}=n/2$. Our one-loop and two-loop results are compared with Petrov's one-loop theory~\cite{petrov2015}. The curves are labeled in the figure. $\mathcal{E}$ is in units of $10^{-4}\hbar^2/(2ma^5)$ and $n$ in units of $10^{-3}a^{-3}$.}
	\label{fig1}
\end{figure}

\begin{figure}
	\centering
	\includegraphics[width=0.45\textwidth]{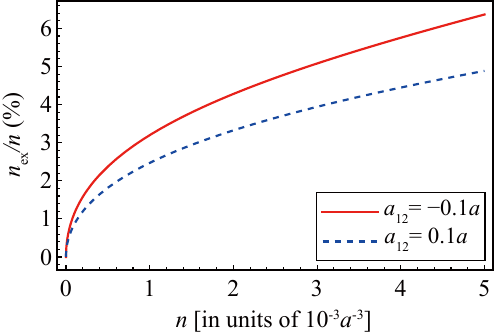}
	\caption{Quantum depletion fraction $n_{\mathrm{ex}}/n$ as a function of the total density $n$ for $a_{12}=-0.1a$ and $a_{12}=0.1a$. Parameters and units are the same as in Fig.~\ref{fig1}.}
	\label{fig2}
\end{figure}

\begin{figure}
	\centering
	\includegraphics[width=0.45\textwidth]{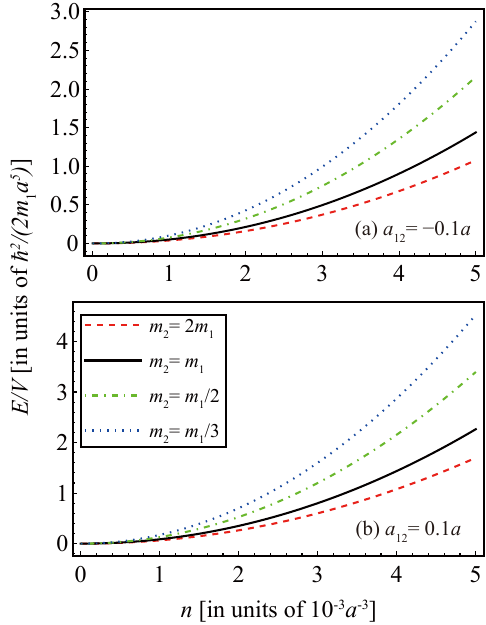}
	\caption{Energy density as a function of the total density $n$ for different mass ratios $r=m_{2}/m_{1}=2,1,1/2$, and $1/3$ (as labeled in the figure), with (a) $a_{12}=-0.1a$ and (b) $a_{12}=0.1a$. Other parameters are as in Fig.~\ref{fig1}. The curves confirm the scaling identity~\eqref{scalingr}: $\mathcal{E}(r)=(1+1/r)\mathcal{E}(1)/2$. $\mathcal{E}$ is in units of $10^{-4}\hbar^2/(2m_{1}a^5)$ and $n$ in units of $10^{-3}a^{-3}$.}
	\label{fig3}
\end{figure}

We first consider the balanced case, i.e., $m_{1}=m_{2}=m$, $a_{11}=a_{22}=a$, and $n_{1}=n_{2}=n/2$. In this situation the self-consistent equations derived in Sec.~\ref{sec:energy} admit closed analytical solutions, and the ground-state energy density takes the compact form
\begin{widetext}
	\begin{equation}
		\mathcal{E}=\frac{g_{12}}{2}n^2+\frac{\dg n^2}{4}\left[1+\frac{128}{15\sqrt{\pi}}\sqrt{\frac{n}{2}(a-a_{12})^3}-\frac{1024}{9\pi}\frac{n}{2}(a-a_{12})^3+\mathcal{O}\left(\left(\frac{n}{2}(a-a_{12})^3\right)^{3/2}\right)\right], \label{ed}
	\end{equation}
\end{widetext}
where $\dg\equiv g-g_{12}=4\pi\hbar^2(a-a_{12})/m$ and $n=n_{1}+n_{2}$ is the total density. Equation~\eqref{ed} is one of the central results of this work, and its structure admits a transparent physical interpretation. The first term is the interspecies mean-field energy. The second term is precisely the sum of the two-loop EOSs of two \emph{independent} single-component Bose gases, each with density $n/2$ and with the intraspecies scattering length replaced by the effective value $\delta a\equiv a-a_{12}$. Indeed, for a single-component gas the two-loop EOS reads $\mathcal{E}_{\mathrm{s}}=(g_{\mathrm{s}}\rho^2/2)[1+\frac{128}{15\sqrt{\pi}}\sqrt{\rho a_{\mathrm{s}}^3}-\frac{1024}{9\pi}\rho a_{\mathrm{s}}^3+\cdots]$~\cite{Song_2022}, which Eq.~\eqref{ed} reproduces species by species under the mapping $g_{\mathrm{s}}\to\dg$, $\rho\to n/2$, $a_{\mathrm{s}}\to a-a_{12}$. In the limit $a_{12}/a\to0$ Eq.~\eqref{ed} therefore reduces exactly to the two-loop EOS of two independent single-component gases, a nontrivial consistency check of our Hubbard-Stratonovich saddle-point scheme. The interspecies interaction enters the universal quantum corrections \emph{exclusively} through the combination $\delta a=a-a_{12}$: attraction ($a_{12}<0$) strengthens the effective scattering length of each species and enhances quantum fluctuations, whereas repulsion ($a_{12}>0$) weakens them. We term this the ``single-component mapping'' of the two-body-correlation universal quantum effect in a weakly coupled mixture. Physically, the mapping originates from the saddle-point treatment, which converts the interspecies interaction into a static (Hartree) background for each species: it renormalizes the effective intraspecies coupling to $\delta g_{\sigma}$ but leaves the fluctuation structure of each sector unchanged. We stress that the next-to-LHY correction in Eq.~\eqref{ed} is a pure two-body-correlation effect: the logarithmic three-body correction $\propto n^{2}(na^{3})\ln(na^{3})$ of a single-component gas~\cite{wu1959,braaten} arises from the setting-sun skeleton diagrams, which are neglected in our two-loop approximation (only the double-bubble diagrams are retained), so that the mapping is exact within this truncation. The same conclusions hold equally for repulsive interspecies interactions ($a_{12}>0$), as demonstrated by the (b) panels of Figs.~\ref{fig1} and \ref{fig3} and by the corresponding curve in Fig.~\ref{fig2}.

The corresponding chemical potential is obtained from $\mu=\partial\mathcal{E}/\partial n$ evaluated along the symmetric configuration $n_{1}=n_{2}=n/2$ (equivalently, $\mu=\mu_{1}=\mu_{2}$), which yields
\begin{widetext}
	\begin{equation}
		\mu=g_{12}n+\frac{\dg n}{2}\left[1+\frac{32}{3\sqrt{\pi}}\sqrt{\frac{n}{2}(a-a_{12})^3}-\frac{512}{3\pi}\frac{n}{2}(a-a_{12})^3+\mathcal{O}\left(\left(\frac{n}{2}(a-a_{12})^3\right)^{3/2}\right)\right], \label{mu}
	\end{equation}
\end{widetext}
again in exact correspondence with the single-component universal series~\cite{Song_2022,zhang2024} under the same mapping: both the leading LHY coefficient $32/(3\sqrt{\pi})$ and the next-to-LHY coefficient $512/(3\pi)$ are reproduced species by species. Likewise, substituting the solutions of the self-consistent equations into Eqs.~\eqref{n1} and \eqref{n2}, the quantum depletion fraction is found as
\begin{widetext}
	\begin{equation}
		\frac{n_{\mathrm{ex}}}{n}=\frac{8}{3\sqrt{\pi}}\sqrt{\frac{n}{2}(a-a_{12})^3}\left[1-\frac{16}{\sqrt{\pi}}\sqrt{\frac{n}{2}(a-a_{12})^3}+\mathcal{O}\left(\frac{n}{2}(a-a_{12})^3\right)\right], \label{dep}
	\end{equation}
\end{widetext}
with $n_{\mathrm{ex}}=n_{1\mathrm{ex}}+n_{2\mathrm{ex}}$. Under the same mapping, Eq.~\eqref{dep} reproduces the single-component depletion $\rho_{\mathrm{ex}}/\rho=(8/(3\sqrt{\pi}))\sqrt{\rho a^{3}}[1-(16/\sqrt{\pi})\sqrt{\rho a^{3}}+\cdots]$~\cite{Song_2022,zhang2024}, whose leading term has been verified experimentally in a homogeneous single-component gas~\cite{lopes2017}. Equations~\eqref{ed}--\eqref{dep} form a complete, parameter-free analytical description of the weakly interacting binary Bose mixture at the two-loop level: for given intraspecies scattering lengths and a given (weak) interspecies scattering length, all equilibrium thermodynamic quantities are determined to beyond-LHY order.

In Fig.~\ref{fig1} we compare the particle-density dependence of the energy density obtained from our one-loop and two-loop results with the prediction of Petrov's theory~\cite{petrov2015,huhui} for $a_{12}=-0.1a$ [panel (a)] and $a_{12}=0.1a$ [panel (b)]. Several observations are in order. First, the two-loop curves lie systematically below the one-loop curves, reflecting the negative beyond-LHY coefficient $-1024/(9\pi)$ in Eq.~\eqref{ed}. At the largest density shown, $na^3\sim8\times10^{-3}$, the two-loop correction reaches about $55\%$ of the one-loop (LHY) correction for $a_{12}=-0.1a$, i.e., it is as important as in the single-component case. Second, our one-loop result agrees with Petrov's theory in the limit $a_{12}/a\to0$ and deviates from it by corrections of order $a_{12}/a$ for finite interspecies coupling, here $|a_{12}|/a=0.1$. This deviation is the price of the saddle-point treatment, which treats the fluctuations of each species as independent single-component fluctuations and neglects the mixed interspecies fluctuation modes. This neglect is systematically controlled by the small parameter $|a_{12}|/a$ and is precisely the regime our approach is designed for. Third, the deviation between the one-loop and two-loop results grows with density and is asymmetric between attraction and repulsion: because the effective gas parameter $\sqrt{(n/2)(a-a_{12})^3}$ is larger (smaller) than $\sqrt{(n/2)a^3}$ for $a_{12}<0$ ($a_{12}>0$), quantum corrections are enhanced by interspecies attraction and suppressed by interspecies repulsion. This asymmetry has no counterpart in a single-component gas and is a genuine mixture effect, directly traceable to the single-component mapping.

The quantum depletion fraction is shown in Fig.~\ref{fig2} for the same parameters. It grows monotonically with density, consistent with Eq.~\eqref{dep}. The next-order correction in Eq.~\eqref{dep} is non-negligible at the highest densities shown, as in the single-component case~\cite{Song_2022}. In line with the single-component mapping, the depletion is enhanced by interspecies attraction ($a_{12}=-0.1a$, for which $\delta a=1.1a$) and suppressed by interspecies repulsion ($a_{12}=0.1a$, $\delta a=0.9a$). Since the condensate fraction is $n_{c}/n=1-n_{\mathrm{ex}}/n$, the measurement of the condensate fraction provides a direct probe of the effective-scattering-length shift $\delta a-a=-a_{12}$ induced by the interspecies interaction, without any additional fitting parameter.

Finally, we turn to mass-imbalance effects, which are unique to heteronuclear mixtures and constitute a distinctive feature of this work. For $m_{1}\neq m_{2}$, keeping $a_{11}=a_{22}=a$ and $n_{1}=n_{2}=n/2$, the mean-field energy depends on the masses only through $g_{12}\propto1/m_{\mathrm{red}}=1/m_{1}+1/m_{2}$ and $g_{\sigma}\propto1/m_{\sigma}$. The fluctuation contributions of species $\sigma$ likewise scale as $m_{\sigma}^{3/2}\delta g_{\sigma}^{5/2}\propto1/m_{\sigma}$ at one-loop order, and at two-loop order the beyond-LHY coefficients depend on the gas parameter $\sqrt{n_{\sigma}(a_{\sigma\sigma}-a_{12})^3}$, which is manifestly mass independent. The entire energy density therefore obeys the exact scaling identity
\begin{equation}
	\mathcal{E}(n;m_{1},m_{2})=\frac{m}{2}\left(\frac{1}{m_{1}}+\frac{1}{m_{2}}\right)\mathcal{E}(n;m,m), \label{scaling}
\end{equation}
where $m$ is an arbitrary reference mass and $\mathcal{E}(n;m,m)$ is the mass-balanced EOS of Eq.~\eqref{ed}. The emergence of the reduced-mass combination $1/m_{1}+1/m_{2}$ mirrors its ubiquitous role in two-body physics, where the relative motion depends on the masses only through the reduced mass $m_{\mathrm{red}}$. Equivalently, in terms of the mass ratio $r\equiv m_{2}/m_{1}$,
\begin{equation}
	\mathcal{E}(n;r)=\frac{1+1/r}{2}\,\mathcal{E}(n;1), \label{scalingr}
\end{equation}
an identity valid already at the one-loop level and persisting at the two-loop level. Although this identity is purely algebraic at the mean-field level, its persistence through the quantum fluctuations is a nontrivial structural property. At one-loop order it relies on the cancellation $m_{\sigma}^{3/2}\delta g_{\sigma}^{5/2}\propto1/m_{\sigma}$ inside the zero-point integrals. At two-loop order, the density-fluctuation integrals $Q_{\sigma}\propto m_{\sigma}^{3/2}M_{\sigma}^{3/2}$ are themselves mass independent, because the self-consistent solution $M_{\sigma}\propto\delta g_{\sigma}n_{\sigma}$ carries the same inverse-mass factor, and every two-loop term inherits the $1/m_{\sigma}$ scaling separately. No mixed mass-ratio combination such as $\sqrt{m_{1}m_{2}}$ or $\ln(m_{1}/m_{2})$ appears, because the saddle-point decoupling renders each species an independent single-component problem whose gas parameter $\sqrt{n_{\sigma}(a_{\sigma\sigma}-a_{12})^{3}}$ is mass independent. A direct treatment based on the mixed $4\times4$ modes would instead introduce mass-imbalance combinations already at the one-loop level. For arbitrary mass ratios, mass imbalance therefore enters the weakly coupled EOS only as a global amplitude, leaving its density dependence unchanged. Moreover, the self-consistently closed nature of the calculation, in which all quantities follow from the same solution of the Schwinger-Dyson equations and thermodynamic consistency is guaranteed by the $\Phi$-derivable structure, makes the scaling law an exact property of the scheme itself rather than an accidental feature of a particular perturbative order. Equations~\eqref{scaling} and \eqref{scalingr} show that, in the symmetric-density setup, mass imbalance amounts to a global rescaling of the EOS, with the lighter species dominating both the mean-field energy and the quantum fluctuations, as both the coupling $g_{\sigma}$ and the zero-point energy of the Bogoliubov modes scale as $1/m_{\sigma}$ at fixed scattering lengths. In Fig.~\ref{fig3} we verify this scaling for the mass ratios $r=2,1,1/2,1/3$: the numerical curves collapse onto Eq.~\eqref{scalingr} for both signs of $a_{12}$.

For realistic heteronuclear mixtures, e.g., $^{41}\mathrm{K}$--$^{87}\mathrm{Rb}$ with $r\simeq87/41\simeq2.12$, the scaling factor is $\simeq0.74$, i.e., a $\sim26\%$ reduction of the energy density with respect to the mass-balanced EOS with the same scattering lengths and densities, a sizable effect that provides a clear benchmark for future numerical studies of heteronuclear mixtures.

Before concluding, we comment on the range of applicability of our results. The parameters of Figs.~\ref{fig1}--\ref{fig3}, $a_{12}=\pm0.1a$ and $na^{3}\lesssim8\times10^{-3}$, lie deep inside the weak-coupling regime $\abs{a_{12}}\ll a$ and the dilute regime $na^{3}\ll1$, where both the saddle-point treatment and the low-density expansion are quantitatively controlled. Two remarks are in order. First, the dilute limit enters only through the analytic expansion of the self-consistent solutions: the two-loop CJT scheme itself is a nonperturbative $\Phi$-derivable resummation, with dressed propagators and the Goldstone theorem preserved by the gapless Hartree-Fock correction, so that the same self-consistent equations can be solved numerically at substantially larger gas parameters. Second, the restriction $\abs{a_{12}}\ll a$ is tied to the saddle-point approximation rather than to the $\Phi$-derivable framework itself, and stronger interspecies couplings can be addressed by including the fluctuations of the auxiliary field. Within the analytic expansion, the expressions remain formally well defined for all $\abs{a_{12}}<a$ (i.e., $\delta g_{\sigma}>0$), away from the droplet boundary $a_{12}\to-a$ and the miscibility boundary $a_{12}\to a$. The accuracy of the saddle-point treatment is quantified at leading order by comparing our one-loop EOS with Petrov's exact one-loop result: the relative deviation of the LHY coefficient, $|2(1-\alpha)^{5/2}-\mathcal{F}(\alpha)|/\mathcal{F}(\alpha)$ with $\alpha=a_{12}/a$, grows from about $25\%$ at $|\alpha|=0.1$ to about $65\%$ at $|\alpha|=0.3$. Since the LHY term contributes only a fraction of the total energy in the dilute regime, this translates into an error of a few percent in the total energy density at $|\alpha|=0.1$ and of order ten percent at $|\alpha|=0.3$. We therefore expect quantitative control for $\abs{a_{12}}/a\lesssim0.1$, as used in the figures, and semiquantitative usefulness up to $\abs{a_{12}}/a\sim0.3$. Within this range the predicted two-loop correction, which reaches about $55\%$ of the LHY term at $na^{3}\sim8\times10^{-3}$, exceeds the estimated systematic error of the saddle-point treatment, so that the beyond-LHY signal is not masked by the approximation error.

\section{Conclusion}
\label{sec:conclusion}
In summary, we have developed a unified and fully analytical theory of the universal quantum corrections to two-body correlations in a weakly interspecies interacting binary Bose mixture at zero temperature. The central strategy is to decouple the interspecies interaction via a Hubbard--Stratonovich transformation and to treat the resulting auxiliary field at the saddle-point level. This procedure reduces the binary mixture to two independent single-component problems with renormalized intraspecies couplings $\delta g_{\sigma}=g_{\sigma}-g_{12}$, which are solved within the two-loop Cornwall--Jackiw--Tomboulis effective action supplemented by the gapless Hartree--Fock correction. The resulting $\Phi$-derivable scheme, formulated in terms of dressed propagators, is self-consistent beyond the plain Bogoliubov expansion and preserves a gapless excitation spectrum. Integrating out the auxiliary field then restores the interspecies mean-field contribution. The approach applies equally to attractive and repulsive interspecies interactions, protects the Goldstone theorem in each sector, and renders the coupled self-consistent problem analytically tractable.

Within this framework, we derived the ground-state energy density, the chemical potential, and the quantum depletion as explicit low-density expansions in powers of the gas parameter. The results reveal a remarkably simple structure, which we call the single-component mapping: the universal quantum corrections of the mixture are obtained by evaluating the single-component universal series at an effective scattering length $\delta a_{\sigma}=a_{\sigma\sigma}-a_{12}$ for each species, while the direct interspecies contribution is entirely contained in the mean-field term. At one-loop order, the equation of state reproduces Petrov's result~\cite{petrov2015,huhui} in the limit $a_{12}/a\to 0$, and at two-loop order it yields the beyond-LHY corrections, which are enhanced by interspecies attraction and suppressed by interspecies repulsion. For heteronuclear mixtures, we further established an exact scaling identity for the energy density under mass imbalance at the one-loop level, which survives the two-loop truncation and captures the mass-ratio dependence of the equation of state through the simple factor $(1+m_{1}/m_{2})/2$. This result is directly relevant to realistic mixtures such as $^{41}$K--$^{87}$Rb.

Our results are universal in that they depend only on the $s$-wave scattering lengths, and they clarify how interspecies coupling renormalizes the two-body-correlation effects of each species. The analytical equation of state derived here also provides a controlled starting point for addressing nonuniversal finite-range effects and three-body correlations. More broadly, multi-component Bose gases continue to attract significant interest, as exemplified by quantum droplets in three-component mixtures~\cite{ma2021,ma2025}. For a fully symmetric $N$-component mixture, in which all species share the same mass, intraspecies coupling, and interspecies coupling, the present scheme extends directly: the single-component mapping remains valid, and the energy density is obtained from Eq.~\eqref{ed} by replacing the per-species density $n/2$ with $n/N$, namely, $\mathcal{E}=g_{12}n^{2}/2+(\delta g\,n^{2}/2N)[1+\frac{128}{15\sqrt{\pi}}\sqrt{(n/N)(a-a_{12})^{3}}-\frac{1024}{9\pi}(n/N)(a-a_{12})^{3}+\cdots]$. 
The chemical potential and the depletion follow from the same substitution. Such closed analytical results for the weakly coupled multi-component case, which follow directly from the single-component mapping, provide useful benchmarks for future studies of multi-component Bose gases. For non-identical species, by contrast, the Hubbard--Stratonovich decoupling requires a matrix of auxiliary fields and the analysis becomes considerably more involved.
We expect that the single-component mapping and the mass-ratio scaling identity derived here will serve as useful benchmarks for experiments and numerical simulations of weakly coupled binary Bose mixtures.

\begin{acknowledgments}
	The authors thank Hui Hu and Yu Jiang for stimulating discussions, and Xiaoran Ye, Yi Zhang, and Ziheng Zhou for useful discussions.
\end{acknowledgments}

\nocite{*}
\bibliography{ref}

@article{Song_2022,
doi = {10.1209/0295-5075/ac8370},
url = {https://doi.org/10.1209/0295-5075/ac8370},
year = {2022},
publisher = {EDP Sciences, IOP Publishing and Societ¨¤ Italiana di Fisica},
volume = {139},
number = {4},
pages = {45001},
author = {Song, Pham The},
title = {Universal quantum effect of two-body correlation in a weakly interacting Bose gas},
journal = {Europhysics Letters}
}

@article{braaten,
  title = {Quantum Corrections to the Energy Density of a Homogeneous {{Bose}} Gas},
  author = {Braaten, E. and Nieto, A.},
  year = {1999},
  journal = {The European Physical Journal B - Condensed Matter and Complex Systems},
  volume = {11},
  number = {1},
  pages = {143--159},
  issn = {1434-6036},
  doi = {10.1007/s100510050925}
}

@article{huhui,
  title = {Consistent Theory of Self-Bound Quantum Droplets with Bosonic Pairing},
  author = {Hu, Hui and Liu, Xia-Ji},
  journal = {Phys. Rev. Lett.},
  volume = {125},
  issue = {19},
  pages = {195302},
  numpages = {6},
  year = {2020},
  month = {Nov},
  publisher = {American Physical Society},
  doi = {10.1103/PhysRevLett.125.195302},
  url = {https://link.aps.org/doi/10.1103/PhysRevLett.125.195302}
}

@article{rmp76.599,
  title = {Theory of the weakly interacting Bose gas},
  author = {Andersen, Jens O.},
  journal = {Rev. Mod. Phys.},
  volume = {76},
  issue = {2},
  pages = {599--639},
  numpages = {0},
  year = {2004},
  month = {Jul},
  publisher = {American Physical Society},
  doi = {10.1103/RevModPhys.76.599},
  url = {https://link.aps.org/doi/10.1103/RevModPhys.76.599}
}

@article{CJT,
  title = {Effective action for composite operators},
  author = {Cornwall, John M. and Jackiw, R. and Tomboulis, E.},
  journal = {Phys. Rev. D},
  volume = {10},
  issue = {8},
  pages = {2428--2445},
  numpages = {0},
  year = {1974},
  publisher = {American Physical Society},
  doi = {10.1103/PhysRevD.10.2428},
  url = {https://link.aps.org/doi/10.1103/PhysRevD.10.2428}
}

@article{dV,
  title = {Gapless {{Hartree-Fock}} resummation scheme for the {{$O(N)$}} model},
  author = {Ivanov, Yu. B. and Riek, F. and Knoll, J.},
  journal = {Phys. Rev. D},
  volume = {71},
  issue = {10},
  pages = {105016},
  numpages = {11},
  year = {2005},
  publisher = {American Physical Society},
  doi = {10.1103/PhysRevD.71.105016},
  url = {https://link.aps.org/doi/10.1103/PhysRevD.71.105016}
}

@book{negele,
author = {Negele, J.W. and Orland, Henri},
title = {Quantum Many-particle Systems},
publisher = {CRC Press},
year = {1998},
url = {https://doi.org/10.1201/9780429497926}
}

@article{bogoliubov1947,
  title = {On the Theory of Superfluidity},
  author = {Bogoliubov, N. N.},
  journal = {J. Phys. (USSR)},
  volume = {11},
  pages = {23},
  year = {1947}
}

@article{lhy1957,
  title = {Eigenvalues and Eigenfunctions of a {Bose} System of Hard Spheres and Its Low-Temperature Properties},
  author = {Lee, T. D. and Huang, K. and Yang, C. N.},
  journal = {Phys. Rev.},
  volume = {106},
  issue = {6},
  pages = {1135--1145},
  numpages = {0},
  year = {1957},
  month = {Jun},
  publisher = {American Physical Society},
  doi = {10.1103/PhysRev.106.1135},
  url = {https://link.aps.org/doi/10.1103/PhysRev.106.1135}
}

@article{wu1959,
  title = {Ground State of a {Bose} System of Hard Spheres},
  author = {Wu, Tai Tsun},
  journal = {Phys. Rev.},
  volume = {115},
  issue = {6},
  pages = {1390--1404},
  numpages = {0},
  year = {1959},
  month = {Sep},
  publisher = {American Physical Society},
  doi = {10.1103/PhysRev.115.1390},
  url = {https://link.aps.org/doi/10.1103/PhysRev.115.1390}
}

@article{petrov2015,
  title = {Quantum Mechanical Stabilization of a Collapsing {Bose-Bose} Mixture},
  author = {Petrov, D. S.},
  journal = {Phys. Rev. Lett.},
  volume = {115},
  issue = {15},
  pages = {155302},
  numpages = {5},
  year = {2015},
  month = {Oct},
  publisher = {American Physical Society},
  doi = {10.1103/PhysRevLett.115.155302},
  url = {https://link.aps.org/doi/10.1103/PhysRevLett.115.155302}
}

@article{cabrera2018,
  title = {Quantum liquid droplets in a mixture of {Bose-Einstein} condensates},
  author = {Cabrera, C. R. and Tanzi, L. and Sanz, J. and Naylor, B. and Thomas, P. and Cheiney, P. and Tarruell, L.},
  journal = {Science},
  volume = {359},
  number = {6373},
  pages = {301--304},
  year = {2018},
  publisher = {American Association for the Advancement of Science},
  doi = {10.1126/science.aao5686},
  url = {https://www.science.org/doi/10.1126/science.aao5686}
}

@article{thalhammer2008,
  title = {Double Species {Bose-Einstein} Condensate with Tunable Interspecies Interactions},
  author = {Thalhammer, G. and Barontini, G. and {De Sarlo}, L. and Catani, J. and Minardi, F. and Inguscio, M.},
  journal = {Phys. Rev. Lett.},
  volume = {100},
  issue = {21},
  pages = {210402},
  numpages = {4},
  year = {2008},
  month = {May},
  publisher = {American Physical Society},
  doi = {10.1103/PhysRevLett.100.210402},
  url = {https://link.aps.org/doi/10.1103/PhysRevLett.100.210402}
}

@article{phat2009,
  title = {Bose--{Einstein} condensation in binary mixture of {Bose} gases},
  author = {Phat, Tran Huu and Hoa, Le Viet and Anh, Nguyen Tuan and Long, Nguyen Van},
  journal = {Annals of Physics},
  volume = {324},
  number = {10},
  pages = {2074--2094},
  year = {2009},
  publisher = {Elsevier},
  doi = {10.1016/j.aop.2009.07.003},
  url = {https://doi.org/10.1016/j.aop.2009.07.003}
}

@article{song2022aop,
  title = {The {Casimir} effect of dual weakly interacting {Bose} gases at zero-temperature},
  author = {Song, P. T.},
  journal = {Annals of Physics},
  volume = {447},
  pages = {169144},
  year = {2022},
  publisher = {Elsevier},
  doi = {10.1016/j.aop.2022.169144},
  url = {https://doi.org/10.1016/j.aop.2022.169144}
}

@article{rakhimov2022,
  title = {Self-consistent theory of a homogeneous binary {Bose} mixture with strong repulsive interspecies interaction},
  author = {Rakhimov, Abdulla and Abdurakhmonov, Tolibjon and Narzikulov, Zabardast and Yukalov, Vyacheslav I.},
  journal = {Phys. Rev. A},
  volume = {106},
  issue = {3},
  pages = {033301},
  numpages = {10},
  year = {2022},
  month = {Sep},
  publisher = {American Physical Society},
  doi = {10.1103/PhysRevA.106.033301},
  url = {https://link.aps.org/doi/10.1103/PhysRevA.106.033301}
}

@article{zhang2024,
  title = {Cornwall-{Jackiw-Tomboulis} effective field theory and the nonuniversal equation of state of an ultracold {Bose} gas},
  author = {Zhang, Yi and Liang, Zhaoxin},
  journal = {Phys. Rev. A},
  volume = {110},
  issue = {4},
  pages = {043318},
  numpages = {18},
  year = {2024},
  month = {Oct},
  publisher = {American Physical Society},
  doi = {10.1103/PhysRevA.110.043318},
  url = {https://link.aps.org/doi/10.1103/PhysRevA.110.043318}
}

@article{myatt1997,
  title = {Production of Two Overlapping {Bose-Einstein} Condensates by Sympathetic Cooling},
  author = {Myatt, C. J. and Burt, E. A. and Ghrist, R. W. and Cornell, E. A. and Wieman, C. E.},
  journal = {Phys. Rev. Lett.},
  volume = {78},
  issue = {4},
  pages = {586--589},
  numpages = {0},
  year = {1997},
  month = {Jan},
  publisher = {American Physical Society},
  doi = {10.1103/PhysRevLett.78.586},
  url = {https://link.aps.org/doi/10.1103/PhysRevLett.78.586}
}

@article{ho1996,
  title = {Binary Mixtures of {Bose} Condensates of Alkali Atoms},
  author = {Ho, Tin-Lun and Shenoy, V. B.},
  journal = {Phys. Rev. Lett.},
  volume = {77},
  issue = {16},
  pages = {3276--3279},
  numpages = {0},
  year = {1996},
  month = {Oct},
  publisher = {American Physical Society},
  doi = {10.1103/PhysRevLett.77.3276},
  url = {https://link.aps.org/doi/10.1103/PhysRevLett.77.3276}
}

@article{hugenholtz1959,
  title = {Ground-State Energy and Excitation Spectrum of a System of Interacting Bosons},
  author = {Hugenholtz, N. M. and Pines, D.},
  journal = {Phys. Rev.},
  volume = {116},
  issue = {3},
  pages = {489--506},
  numpages = {0},
  year = {1959},
  month = {Nov},
  publisher = {American Physical Society},
  doi = {10.1103/PhysRev.116.489},
  url = {https://link.aps.org/doi/10.1103/PhysRev.116.489}
}

@article{derrico2019,
  title = {Observation of quantum droplets in a heteronuclear bosonic mixture},
  author = {D'Errico, C. and Burchianti, A. and Prevedelli, M. and Salasnich, L. and Ancilotto, F. and Modugno, M. and Minardi, F. and Fort, C.},
  journal = {Phys. Rev. Research},
  volume = {1},
  issue = {3},
  pages = {033155},
  numpages = {5},
  year = {2019},
  month = {Dec},
  publisher = {American Physical Society},
  doi = {10.1103/PhysRevResearch.1.033155},
  url = {https://link.aps.org/doi/10.1103/PhysRevResearch.1.033155}
}

@article{lopes2017,
  title = {Quantum Depletion of a Homogeneous {Bose-Einstein} Condensate},
  author = {Lopes, R. and Eigen, C. and Navon, N. and Cl{\'e}ment, D. and Smith, R. P. and Hadzibabic, Z.},
  journal = {Phys. Rev. Lett.},
  volume = {119},
  issue = {19},
  pages = {190404},
  numpages = {5},
  year = {2017},
  month = {Nov},
  publisher = {American Physical Society},
  doi = {10.1103/PhysRevLett.119.190404},
  url = {https://link.aps.org/doi/10.1103/PhysRevLett.119.190404}
}

@book{stoof,
  title = {Ultracold {{Quantum Fields}}},
  author = {Stoof, Henk T. C. and Dickerscheid, Dennis B. M. and Gubbels, Koos},
  year = {2009},
  publisher = {Springer},
  isbn = {978-1-4020-8762-2}
}

@book{nagaosa,
  title = {Quantum {{Field Theory}} in {{Condensed Matter Physics}}},
  author = {Nagaosa, Naoto},
  year = {1999},
  publisher = {Springer},
  isbn = {978-3-540-65537-4}
}

@book{Zhai_2021,
 title={Ultracold Atomic Physics},
 publisher={Cambridge University Press},
 author={Zhai, Hui},
 year={2021}
}

@article{ma2021,
  title = {Borromean Droplet in Three-Component Ultracold {Bose} Gases},
  author = {Ma, Yong-Chuan and Peng, Shi-Guo and Cui, Xiaoling},
  journal = {Phys. Rev. Lett.},
  volume = {127},
  issue = {4},
  pages = {043002},
  numpages = {6},
  year = {2021},
  month = {Jul},
  publisher = {American Physical Society},
  doi = {10.1103/PhysRevLett.127.043002},
  url = {https://link.aps.org/doi/10.1103/PhysRevLett.127.043002}
}

@article{ma2025,
  title = {Shell-Shaped Quantum Droplet in a Three-Component Ultracold {Bose} Gas},
  author = {Ma, Yong-Chuan and Cui, Xiaoling},
  journal = {Phys. Rev. Lett.},
  volume = {134},
  issue = {4},
  pages = {043402},
  numpages = {7},
  year = {2025},
  month = {Jan},
  publisher = {American Physical Society},
  doi = {10.1103/PhysRevLett.134.043402},
  url = {https://link.aps.org/doi/10.1103/PhysRevLett.134.043402}
}

@CONTROL{apsrev42Control,
    author="08",
    editor="1",
    pages="0",
    title="0",
    year="1",
    eprint="1"
}

@article{luttinger1960,
  title = {Ground-State Energy of a Many-Fermion System. {II}},
  author = {Luttinger, J. M. and Ward, J. C.},
  journal = {Phys. Rev.},
  volume = {118},
  issue = {5},
  pages = {1417--1427},
  numpages = {0},
  year = {1960},
  month = {Jun},
  publisher = {American Physical Society},
  doi = {10.1103/PhysRev.118.1417},
  url = {https://link.aps.org/doi/10.1103/PhysRev.118.1417}
}

@article{tan2008,
  title = {Three-boson problem at low energy and implications for dilute {Bose-Einstein} condensates},
  author = {Tan, Shina},
  journal = {Phys. Rev. A},
  volume = {78},
  issue = {1},
  pages = {013636},
  numpages = {7},
  year = {2008},
  publisher = {American Physical Society},
  doi = {10.1103/PhysRevA.78.013636}
}

\end{document}